\documentclass[twocolumn,twocolappendix]{aastex63}

\usepackage{bm}
\usepackage{url}
\usepackage{amsmath}
\usepackage{color}
\usepackage{graphicx}

\received{}
\revised{}
\accepted{}
\submitjournal{ApJ}

\shorttitle{ngVLA}
\shortauthors{Ueda et al.}

\begin{document}

\title{
Probing Inner-Edge of Dead Zones in Protoplanetary Disks with ALMA and Next Generation Very Large Array
}

\correspondingauthor{Takahiro Ueda}
\email{takahiro.ueda@nao.ac.jp}

\author[0000-0003-4902-222X]{Takahiro Ueda}
\affil{National Astronomical Observatory of Japan, Osawa 2-21-1, Mitaka, Tokyo 181-8588, Japan}

\author{Luca Ricci}
\affil{Department of Physics and Astronomy, California State University Northridge, 18111 Nordhoff Street, Northridge, CA 91330, USA}

\author[0000-0002-9298-3029]{Mario Flock}
\affil{Max-Planck Institute for Astronomy, K\"{o}nigstuhl 17, 69117 Heidelberg, Germany}

\author{Zachary Castro}
\affil{Department of Physics and Astronomy, California State University Northridge, 18111 Nordhoff Street, Northridge, CA 91330, USA}



\begin{abstract}
The discovery of substructures in protoplanetary disks with ALMA has provided us key insights on the formation of planets.
However, observational constraints on the formation of rocky planets have been still sparse, especially because of the limited spatial resolution. 
The inner edge of so-called dead zone is one of the preferential sites of rocky planet formation.
We investigate the capabilities of ALMA and ngVLA for observing a dust concentration expected at the inner edge of the dead-zone around a Herbig star.
Herbig Ae/Be stars are useful laboratories for exploring the evolution of rocky grains in protoplanetary disks because of their high luminosity which pushes the dead-zone inner edge outward.
We find that, thanks to its unprecedented angular resolution and sensitivity, ngVLA can detect the dust concentration at the dead-zone inner edge, with a reasonable integration time of 10 hrs at $\lambda=3,7$ mm and 1 cm.
The dust concentration is expected to be optically thick at the ALMA wavelengths and cannot be spatially resolved due to its limited resolution.
On the other hand, the flux density from the inner disk regions ($\sim$3--4 au) observed with current VLA is higher for disks with a dust ring, and hence would be a useful indicator that help us choose potential candidates of disks having a dust concentration at the inner most region.
With these observations we can characterize the process of dust concentration in the innermost disk regions, where rocky planets can form.
\end{abstract}


\section{Introduction} \label{sec:intro}
Understanding the formation process of rocky planets is of great importance in planetary science. 
One preferential site of rocky planetesimal/planet formation is the inner edge of the so-called dead zone (\citealt{Kretke2009,Tan2015,Ueda+19,Flock2019,Jankovic2021,Ueda+21}).
The dead zone is the location where magneto-rotational instability (MRI, \citealt{Balbus1998}) is suppressed because of poor gas ionization \citep{Gammie1996}.
The dead zone is likely to have an inner edge where the gas temperature $T$ reaches $\sim$800--1000~K, above which thermal ionization of the gas is effective enough to activate MRI \citep{Gammie1996,Desch2015}.
Across the dead-zone inner edge, the turbulent viscosity arising from MRI steeply decreases from inside out, resulting in a local maximum in the radial profile of the gas pressure (e.g., \citealt{Dzyurkevich2010}; \citealt{Lyra2012}; \citealt{Flock2016}; \citealt{Flock2017}).
The pressure maximum traps solid particles \citep{Whipple1972, Adachi1976} and the local dust-to-gas mass ratio increases, leading potentially to the formation of rocky planetesimals via the streaming instability \citep{Youdin2005, Johansen2007, Carrera2015, Abod2019} or via the gravitational instability \citep{CMF1981}.

It is definitely important to observationally probe whether dust grains indeed pileup at the dead-zone inner edge. Until now, most of the observational constraints from the inner disk are derived from the near infrared (NIR) emission. First observations of young Herbig Ae/Be stars found a NIR excess, which could not be explained by simple radiative disk models \citep{Hillenbrand1992}. Since then, the theoretical models of the inner rim advanced, including the effect of the inner rim shape \citep{Dullemond+01}, dust halo \citep{Vinkovic2006}, grain sizes \citep{Kama2009} and hydrodynamics \citep{Flock2016}. However, the near infrared emission allows us only to probe the very hot (above 1000~K) and low density regions of the uppermost disk atmosphere \citep{Lazareff2017,gravity2019}. To probe the planet formation in the disk midplane we need high angular resolution observations at longer wavelengths.

In the past years, observations with the Atacama Large Millimeter/submillimeter Array (ALMA) have shown that ring and gap structures are very common in protoplanetary disks (e.g., \citealt{ALMA2015,Andrews2018,Andrews20}). 
However, most of observed rings are located far from the central star (stellocentric radii $>$ 10--20 au) and are thought to be composed of icy grains. 
In our work we will for the first time model and present predictions for the dust concentration regions at the inner disk close to the dead-zone edge.
In our work we will explore the potential of current and future sub-mm and radio interferometers, such as ALMA and the Next Generation Very Large Array \citep[ngVLA,][]{Selina2018}, to observe the dust emission from these regions in protoplanetary disks.

\section{Models and Methods} \label{sec:analysis}

\subsection{Dust disk}
In this work, we use the surface density model obtained by \citet{Ueda+19} which investigated the evolution of gas and dust in a disk around a Herbig star.
The central star has a mass of $2.5~M_{\odot}$, radius of $2.5~R_{\odot}$ and effective temperature of 10000 K, resulting in a luminosity of $56~L_{\odot}$.
Our model parameters are similar to those of, e.g., HD179218 and BF Ori (63.1 and 61.6$L_{\odot}$, respectively; \citealt{Mendigutia+11}).
In the simulations, the evolution of rocky grains in a gas disk with a dead-zone is computed by taking into account the collisional growth/fragmentation, turbulent diffusion and radial drift.
The dust-size evolution is computed with a single-size approach in which we compute the evolution of grains with a representative size, i.e., maximum size, in each radial grid cell \citep{Sato+16}.
As the initial condition, the dust surface density is set to be 100 times lower than the gas surface density which is determined by the assumption of a steady state accretion with an accretion rate of $10^{-8}~ M_{\odot}~\rm{yr}^{-1}$. 
In this work we use snapshots at $t=3\times10^{5}$ yr of the simulations.
We refer the readers to \citet{Ueda+19} for the details of the simulations.

\begin{figure}[ht]
\begin{center}
\includegraphics[scale=0.65]{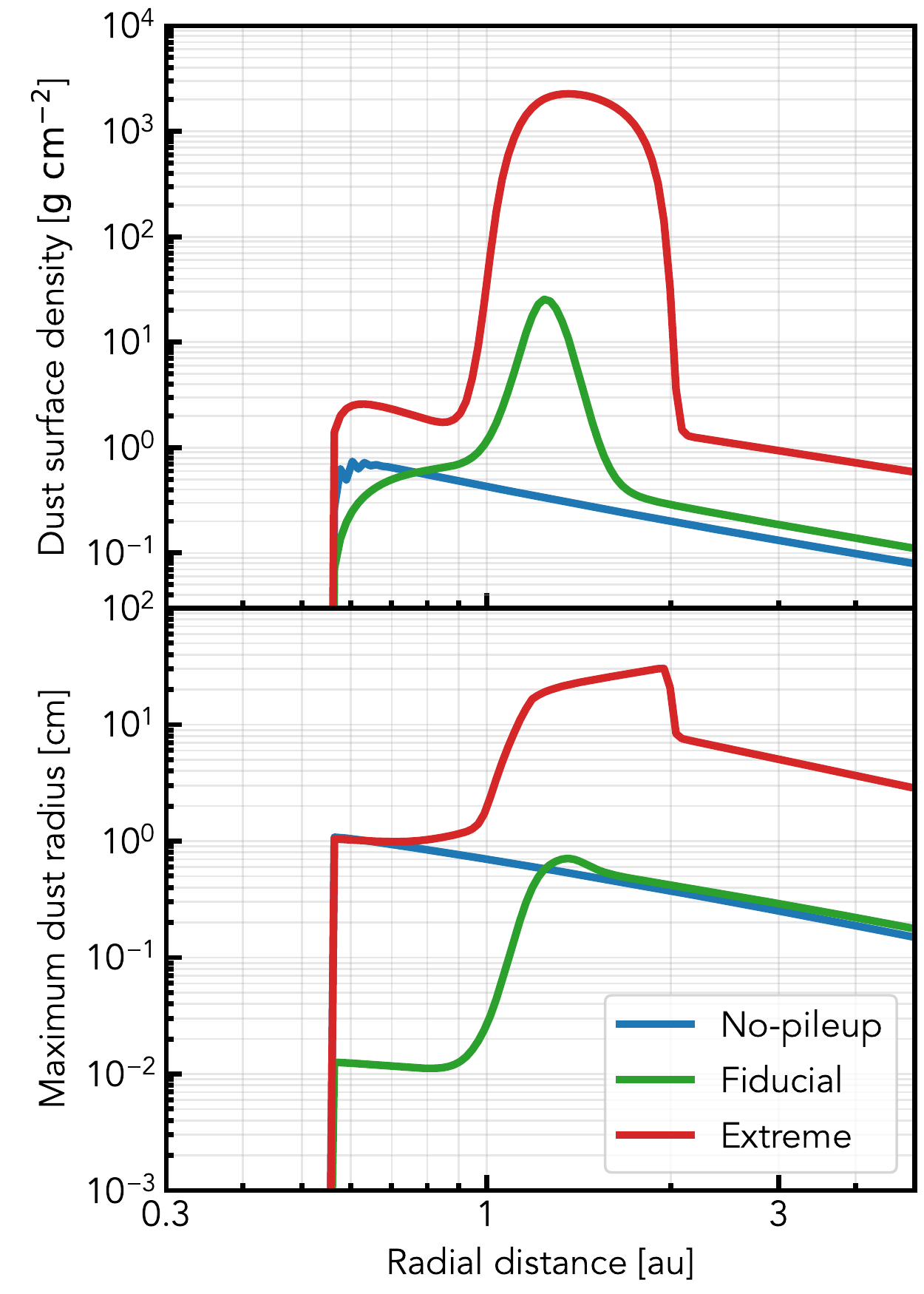}
\caption{
Dust surface density and dust size profile around the dead-zone inner edge.
}
\label{fig:setup}
\end{center}
\end{figure}

We consider three models for the dust distribution named ``No-pileup'', ``Fiducial'' and ``Extreme'', as shown in Figure \ref{fig:setup}.
The No-pileup model corresponds to the disk with no dead-zone inner edge. 
In this model, the turbulence parameter $\alpha_{\rm turb}$ is set to be $10^{-3}$ and is uniform in the entire region.
Because of the uniform turbulence, the No-pileup model has a smooth radial density profile with no dust concentration.
The dust surface density steeply decreases at 0.55 au where rocky grains evaporate into the gas phase.
The Fiducial model has $\alpha_{\rm turb}=10^{-2}$ and $10^{-3}$ in the MRI-active and MRI-dead regions, respectively, with a steep transition at the dead-zone inner edge located at $\sim$ 1 au.
In this model, due to the relatively strong turbulence, the dust size is regulated to $<1$ cm by the turbulence-induced collisional fragmentation.
\begin{figure*}[t]
\begin{center}
\includegraphics[scale=0.44]{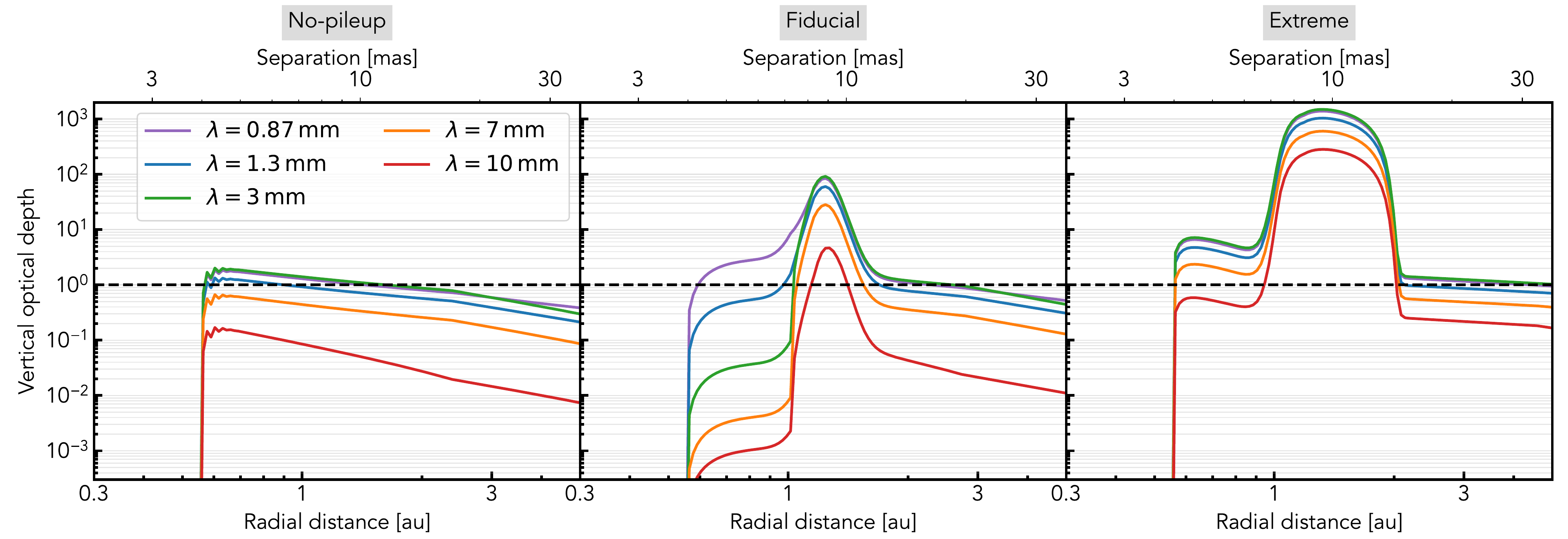}
\caption{
Extinction optical depths in the vertical direction for a face-one oriented disk models at ALMA and ngVLA wavelengths.
The lower x-axis shows the radial distance and the upper x-axis corresponds to the angular separation with a distance of 140 pc. 
}
\label{fig:tau}
\end{center}
\end{figure*}
Owing to their small size, dust particles are marginally coupled with the gas and hence the turbulent diffusion prevents dust from accumulating efficiently.
The midplane dust-to-gas mass ratio is $\sim$ 3, which would trigger the formation of rocky planetesimals via the streaming instability.
The Extreme model is an extreme case where the dust particles strongly pile up at the dead-zone inner edge.
This model has $\alpha_{\rm turb}=10^{-3}$ and $10^{-4}$ in the MRI-active and MRI-dead regions, respectively.
In this model, due to the weak turbulence, turbulent diffusion no longer prevents dust from trapping at the dead-zone inner edge.
In all models, the critical fragmentation velocity of rocky grains is set to 10 ${\rm m~s^{-1}}$.
The choice of the fragmentation velocity is based on the theoretical models of collisions of silicate aggregates composed of $\sim$ 0.1 ${\rm \mu m}$ monomers (e.g., \citealt{Wada+13}).
The exact value of the fragmentation velocity depends on the composition, monomer size and surface energy, and has large uncertainty (e.g., \citealt{BW08,Homma+19,Steinpilz+19}).
With these three models we cover the most extreme cases, from having no pressure bump and no dust concentration until having a pressure bump with very efficient dust trapping.

\subsection{Radiative transfer simulation}
In order to obtain predictions for the maps of the dust continuum emission for the models described above, we perform the three-dimensional radiative transfer simulations with the Monte Carlo radiative transfer code RADMC-3D \citep{RADMC}.
At each radial grid cell, the vertically integrated grain size distribution is assumed to be a truncated power-law distribution ranging from 0.1 ${\rm \mu m}$ to $a_{\rm max}$ with the power-law index of 3.5.
The maximum grain radius $a_{\rm max}$ is determined by the simulation and shown in Figure \ref{fig:setup}.
In the vertical direction, we assume a mixing-settling equilibrium for each dust size \citep{YL2007}; larger grains settle more to the mid-plane.
The grain size distribution is divided into 7 bins, $<0.3~{\rm \mu m}$, 0.3--3 ${\rm \mu m}$, 3--30 ${\rm \mu m}$, 30--300 ${\rm \mu m}$, 0.3--3 ${\rm mm}$, 3--30 ${\rm mm}$ and 3--30 ${\rm cm}$.
We adopt a representative size for each size-bin ($0.1~{\rm \mu m}$, 1 ${\rm \mu m}$, 10 ${\rm \mu m}$, 100 ${\rm \mu m}$, 1 ${\rm mm}$, 10 ${\rm mm}$ and 10 ${\rm cm}$ for bin of $<0.3~{\rm \mu m}$, 0.3--3 ${\rm \mu m}$, 3--30 ${\rm \mu m}$, 30--300 ${\rm \mu m}$, 0.3--3 ${\rm mm}$, 3--30 ${\rm mm}$ and 3--30 ${\rm cm}$, respectively), and compute a spatial distribution and opacities of each size-bin using the representative size.
The opacities are computed using Mie theory for each size bin assuming spherical compact grains with the material density of 3.0 ${\rm g~cm^{-3}}$ and the optical constants of amorphous silicate \citep{Jaeger1994}.
We first perform the iterative thermal Monte Carlo simulations and obtain the temperature structure of the disk, and then perform the imaging simulation using the obtained temperature structure (see \citealt{Ueda+19}).

Figure \ref{fig:tau} shows vertical extinction optical depths of our disk models at ALMA and ngVLA wavelengths.
The optical depth of the No-pileup model has a smooth radial profile and is close to unity at ALMA wavelengths and lower than unity at ngVLA wavelengths.
The Fiducial model has a peak in the optical depth profile at $\sim$ 1 au.
The optical depth at the peak reaches $\sim100$ at $\lambda\lesssim3$ mm, while it is down to $\sim3$ at ngVLA Band 4 ($\lambda=1.0$ cm).
In the Extreme model, the dust ring is quite optically thick ($>$100) and extends to $\sim$ 2 au.

\subsection{Simulations of ALMA and ngVLA observations}
The synthetic images for the dust continuum emission obtained in the previous section were used to simulate the results of observations with ALMA and ngVLA. 
This is done using the \texttt{CASA} software package\footnote{\texttt{https://casa.nrao.edu/}}, following the method outlined in \citet{Ricci2018} \citep[see also][]{Harter2020,Blanco2021}. 
Synthetic observations with ALMA were produced at wavelengths of 870 ${\rm \mu m}$ and 1.3 mm, whereas for the ngVLA we focused on wavelengths of 3 and 7 mm, and 1 cm. In order to provide a fair comparison between the imaging capabilities of these two facilities, the Declination sky coordinate of the disk center was set to $+24~\deg$ (Taurus region) and to $-24~\deg$ (Ophiuchus region) for the ngVLA and ALMA simulations, respectively. 
At the latitudes of these observatories, these correspond to maximum source elevations of about 80 and 89 degrees for the ngVLA and ALMA, respectively.

The model synthetic images are Fourier transformed and sampled at the $(u,v)$ points of the ALMA and ngVLA observations using the \texttt{SIMOBSERVE} task in \texttt{CASA}. 
For ALMA, we considered the antenna position file \texttt{alma.out28.cfg} available in the \texttt{CASA} software package, which contains the longest baselines for this array, and simulated aperture synthesis for an integration time of 10 hours.
For the ngVLA, we consider the \texttt{REV-B} array configuration, which contains 214 antennas of 18-meter diameter. 
The interferometric visibilities are manually corrupted using the \texttt{SETNOISE} task. 
The rms noise values obtained on the maps correspond to an integration time of 10 hours according to the ngVLA array design specified in \citet{Selina2018}.  

The imaging of the interferometric visibilities is performed using the \texttt{TCLEAN} task. 
For both the ALMA and ngVLA datasets we adopt a Briggs weighting scheme with robust parameter $R=-1$ and \texttt{multiscale} deconvolver. 
These choices provide a good trade-off between angular resolution and sensitivity for the emission scales and brightness of the disk models considered in this work~\citep[e.g.,][]{Blanco2021}. For example, imaging with robust parameters $R > -1$ provides better signal-to-noise ratios but at the expense of angular resolution and some of the substructures presented in this work are not spatially resolved; conversely, imaging with robust parameters $R < -1$ provides better angular resolution, but the lower signal-to-noise for the surface brightness does not allow us to detect many of the fainter disk substructures predicted by our models.
The distance to all our disk models was set to 140 pc, which is close to the average distance of several nearby star forming regions.

\section{Results} \label{sec:results}

\subsection{Synthetic images and radial intensity profile}
\begin{figure*}[ht]
\begin{center}
\includegraphics[scale=0.44]{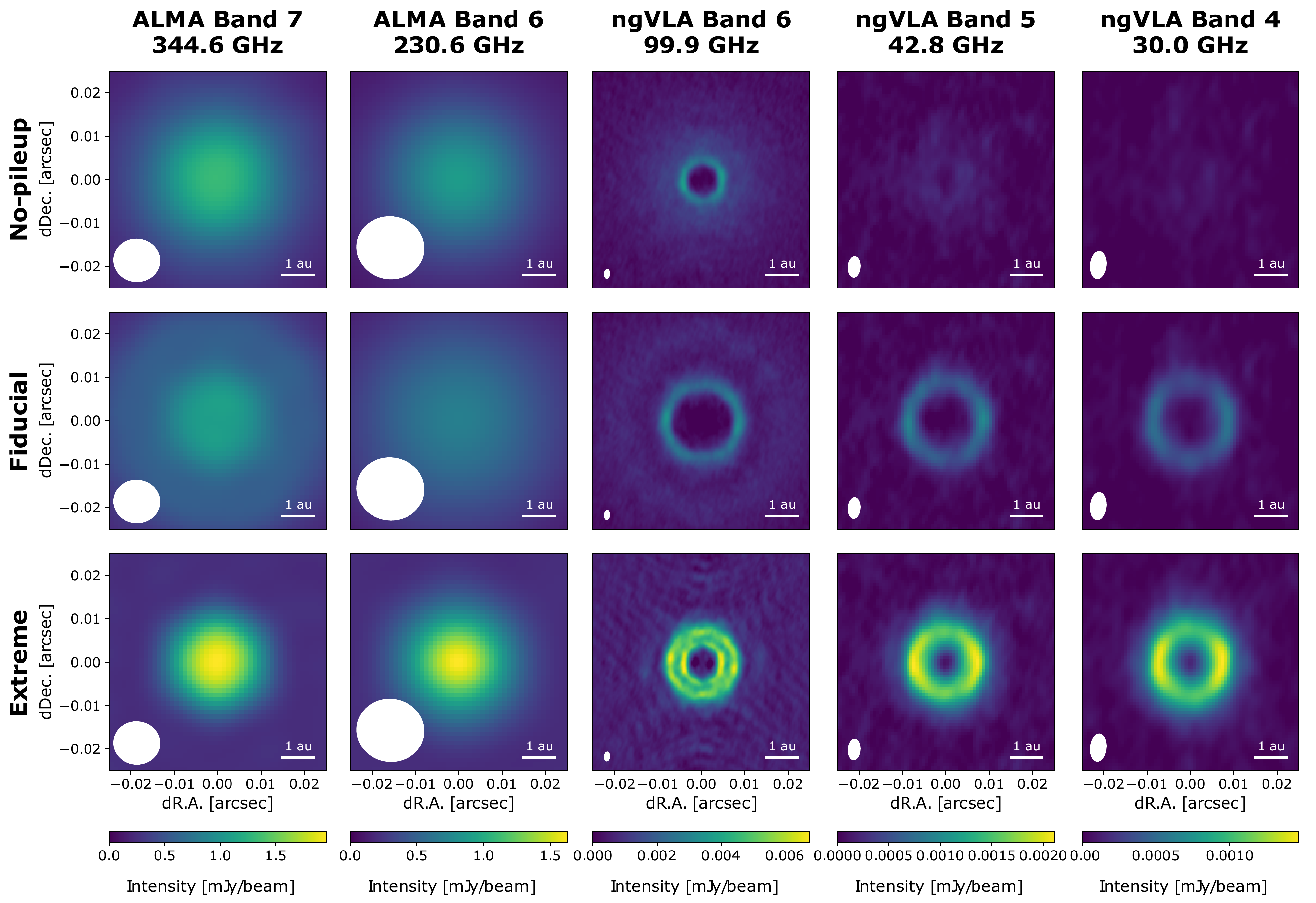}
\caption{
Synthetic images of the innermost region of disks observed with ALMA and ngVLA.
Each row corresponds to the No-pileup (top), Fiducial (middle) and Extreme (bottom) model.
The integration time is set to 10 hrs.
The observing beam-size is denoted with a white-filled ellipse at lower left in each panel.
}
\label{fig:amorphous}
\end{center}
\end{figure*}
\begin{figure*}[ht]
\begin{center}
\includegraphics[scale=0.51]{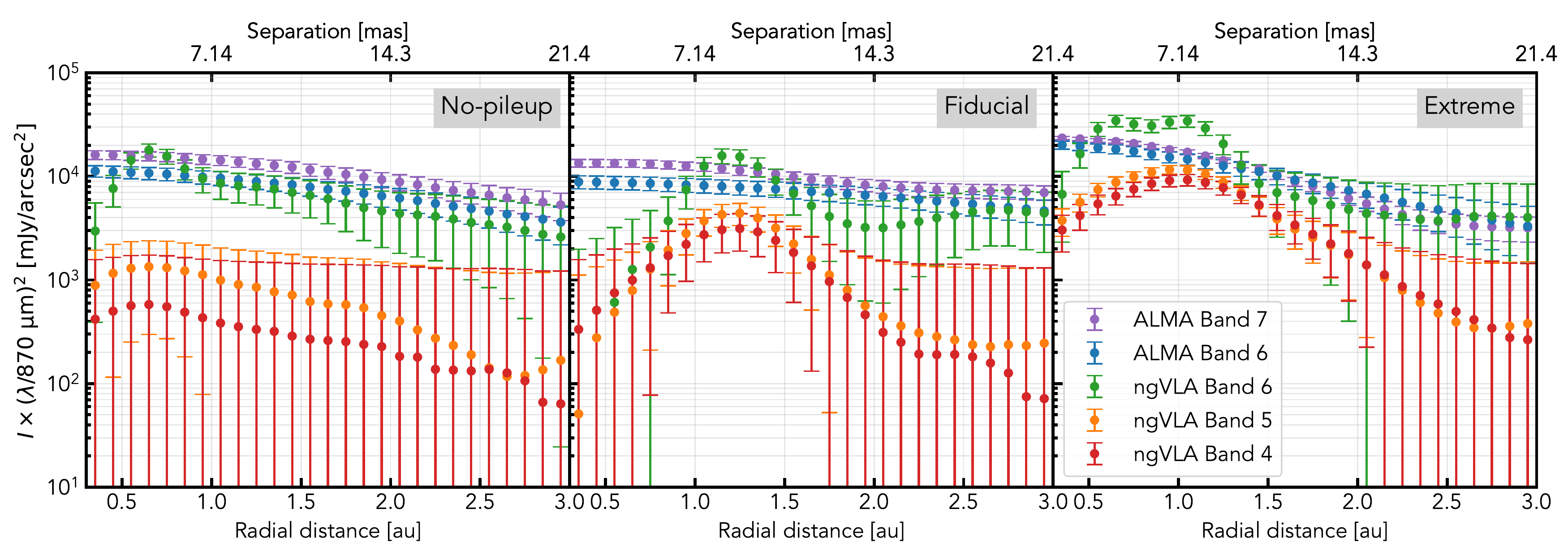}
\caption{
Azimuthally-averaged radial intensity profile of the No-pileup model (left), Fiducial model (center) and Extreme model (right) observed at ALMA and ngVLA wavelengths.
The radial bins are sampled at a width of 0.1 au.
The error bar denotes the standard deviation of the intensities divided by the number of beam segments at each radial bin .
The intensity is multiplied by a square of the wavelength to compensate for the spectral dependence of the Black-body radiation in Rayleigh-jeans limit.
}
\label{fig:radial}
\end{center}
\end{figure*}

Figure \ref{fig:amorphous} shows the synthetic images of the inner regions of the disk models considered here, as observed with ALMA and ngVLA.
The angular resolutions of these synthetic observations are: $10.6\times9.74$ mas (ALMA at 0.87 mm), $15.4\times14.3$ mas (ALMA at 1.3 mm), $2.04\times1.14$ mas (ngVLA at 3 mm), $4.75\times2.66$ mas (ngVLA at 7 mm), $6.30\times3.51$ mas (ngVLA at 1 cm).
The corresponding radial profiles of the azimuthally averaged intensity 
are shown in Figure \ref{fig:radial}.
The peak intensity, rms noise and peak signal-to-noise ratio (SNR) in each map are listed in Table \ref{table:0}.

As a result of the limited angular resolution, the ALMA Band 6 and 7 observations show smooth intensity images with no clear signature of the dust concentration at the dead-zone inner edge.
In the No-pileup model, a bright ring at $r\sim0.5$ au, corresponding to the inner rim of the dust disk, is observed at ngVLA Band 6 with a SNR $\approx~50$.
In ngVLA Band 4 and 5 images, however, the dust rim is not observed clearly.
This is because the disk is optically thin so that the rim-structure is not detected with sufficient SNR at these wavelengths even after 10 hrs of integration time.

In the Fiducial and Extreme models, on the other hand, the ngVLA maps a bright ring at $\sim$ 1 au is clearly detected, which is caused by the dust-pileup at the dead-zone inner edge. 
In the Fiducial model, the ring is detected with SNR ratios of 45.4, 29.3 and 27.5 at ngVLA Band 6, 5 and 4, respectively.
In the Extreme model, the ring is detected with SNR ratios of 82.3, 61.7 and 70.7 at ngVLA Band 6, 5 and 4, respectively.
These high SNR ratios indicate that the dust ring at the dead-zone inner edge can be detected even with a significantly shorter integration time.
If we set the integration time to 3 hrs, the SNR would decrease by a factor of $\sqrt{3/10}\sim0.5$, which is still enough to detect the ring.
In the Extreme model, both rings at the rim of the dust disk and at the dead-zone inner edge are detected at ngVLA Band 6, while they are not resolved at ngVLA Band 4 and 5 due to the poorer resolution.

\begin{table}[ht]
\begin{center}
\caption{Peak intensity, rms noise and SNR of the dust emission in the synthetic images.}
\begin{tabular}{ccccc}
  \hline  
                & $\lambda$ & Peak & Noise & SNR \\
                & [mm]       & [mJy/beam] & [mJy/beam] &                 \\ \hline  \hline
                & 0.87       & 1.32                & $7.68\times10^{-3}$ & 172 \\
                & 1.3        & $8.89\times10^{-1}$ & $4.48\times10^{-3}$ & 198 \\
  No-pileup     & 3.0        & $3.75\times10^{-3}$ & $7.45\times10^{-5}$ & 50.3 \\
                & 7.0        & $2.74\times10^{-4}$ & $3.40\times10^{-5}$ & 8.06 \\
                & 10         & $1.32\times10^{-4}$ & $2.06\times10^{-5}$ & 6.41 \\ \hline
                & 0.87       & 1.10                & $8.01\times10^{-3}$ & 137 \\
                & 1.3        & $6.94\times10^{-1}$ & $4.48\times10^{-3}$ & 155 \\
  Fiducial      & 3.0        & $3.49\times10^{-3}$ & $7.68\times10^{-5}$ & 45.4 \\
                & 7.0        & $9.96\times10^{-4}$ & $3.40\times10^{-5}$ & 29.3 \\
                & 10         & $5.66\times10^{-4}$ & $2.06\times10^{-5}$ & 27.5 \\ \hline
                & 0.87       & 1.95                & $7.80\times10^{-3}$ & 250 \\
                & 1.3        & 1.63                & $4.45\times10^{-3}$ & 366 \\
  Extreme       & 3.0        & $6.78\times10^{-3}$ & $8.24\times10^{-5}$ & 82.3 \\
                & 7.0        & $2.11\times10^{-3}$ & $3.42\times10^{-5}$ & 61.7 \\
                & 10         & $1.47\times10^{-3}$ & $2.08\times10^{-5}$ & 70.7 \\
  \hline  
\end{tabular}
\end{center}
\label{table:0}
\end{table}

\subsection{Flux densities from the inner disk regions}
As shown in the previous section, future observations with the ngVLA can provide evidence for the dust concentration expected by models of dust evolution at the dead-zone inner edge.
Here we investigate the possibility to detect a signature of the dust concentration at the dead-zone inner edge with current VLA resolution.
\begin{table}[ht]
\begin{center}
\caption{Flux density coming from inside 3.5 au}
\begin{tabular}{cccc}
  \hline  
                    & No-pileup & Fiducial & Extreme \\
                    & [mJy]     & [mJy]       & [mJy]         \\\hline  \hline
  870~${\rm \mu m}$ & 11.0                & 11.2                & 9.11 \\
  1.3~${\rm mm}$    & 3.37                & 3.54                & 4.04 \\
  3.0~${\rm mm}$    & $5.11\times10^{-1}$ & $5.54\times10^{-1}$ & $8.76\times10^{-1}$ \\
  7.0~${\rm mm}$    & $8.13\times10^{-3}$ & $1.77\times10^{-2}$ & $4.77\times10^{-2}$ \\
  1.0~${\rm cm}$    & $1.67\times10^{-3}$ & $6.29\times10^{-3}$ & $2.06\times10^{-2}$ \\
  \hline  
\end{tabular}
\end{center}
\label{table:1}
\end{table}
Table \ref{table:1} summarizes the flux density coming from the inner 3.5 au region (diameter of 7 au).
The spatial resolution of 7 au corresponds to an angular resolution of $\sim50$ mas with the distance of 140 pc, which is comparable to the highest resolution of current VLA.
At the wavelength of 870 ${\rm \mu m}$, the flux density has no significant difference between the No-pileup and Fiducial model, while the Extreme model predicts lower flux density.
This is because the intensity in the Extreme model steeply drops radially behind the dead-zone inner edge (see Figure \ref{fig:radial}) because the disk temperature is low due to the shadowing by the dust-pileup (\citealt{Ueda+19}, see Appendix \ref{sec:app}).
At wavelengths longer than 1.3 mm, the Extreme model predicts higher flux density than the other models because of higher optical depth.
The Fiducial model has flux density comparable to that of the No-pileup model at $\lambda=0.87, 1.3$ and 3.0 mm.
Interestingly, on the other hand, at $\lambda=7.0$ mm and 1 cm, the Fiducial model predicts significantly higher flux density than the No-pileup model because of the dust-ring at the dead-zone inner edge.
Therefore, VLA observations at $\lambda=7.0$ mm and 1 cm with the highest resolution could be a useful indicator for inferring potential candidates of disks having dust-concentration at the dead-zone inner edge, even though it cannot resolve the ring-structure spatially.
Here we note that the sub-cm/cm emission from disks is often affected by emission from ionized gas via a variety of different mechanisms~\citep[e.g.,][]{Ubach+17,Ricci+21}. 
This contribution can be estimated  using observations at longer wavelength ($>1$ cm) where the emission is dominated by the ionized gas \citep{Carrasco-Gonzalez+16}.

\section{Dead-zone inner edge in known Herbig Ae/Be disks}
In this section, we discuss possible targets for the observation of the dead-zone inner edge.
In passive disks, the location of the dead-zone inner edge $r_{\rm DZIE}$ can be roughly evaluated as (\citealt{Ueda+17}, see also \citealt{Flock2016})
\begin{eqnarray}
r_{\rm DZIE} \sim 0.8\left( \frac{L_{*}}{56~L_{\odot}} \right)^{1/2}~{\rm au}.
\label{eq:dib}
\end{eqnarray}
Since the dead-zone inner edge is characterized as the region where the gas temperature reaches $\sim$ 1000 K, disks around more luminous stars are more suitable for probing the dead-zone inner edge.
Using Equation \eqref{eq:dib}, the expected angular separation between the central star and the dead-zone inner edge $\theta_{\rm DZIE}$  is evaluated as
\begin{eqnarray}
\theta_{\rm DZIE}\sim5.7\left(\frac{L_{*}}{56L_{\odot}}\right)^{1/2}\left(\frac{d}{\rm 140~pc}\right)^{-1} {\rm mas}
\label{eq:dib_theta}
\end{eqnarray}

\begin{figure}[ht]
\begin{center}
\includegraphics[scale=0.33]{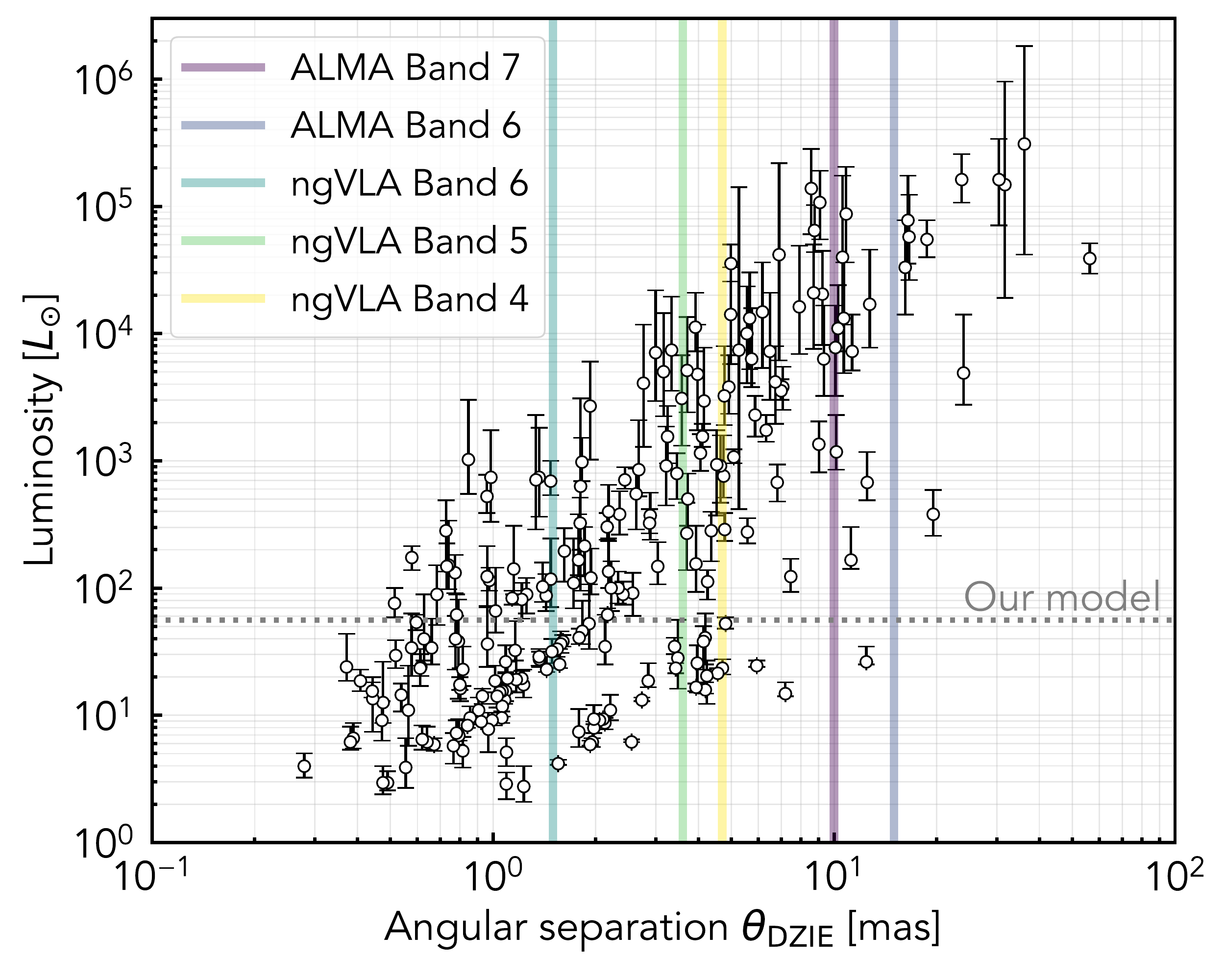}
\caption{
Expected angular separation of the dead-zone inner edge for 218 Herbig disks.
The stellar parameters (luminosity and distance) are taken from \citet{Vioque+18}.
The vertical solid lines show the angular resolutions of ALMA and ngVLA obtained in our synthetic images; 10, 15, 1.5, 3.6 and 4.7 mas for ALMA Band 7, ALMA Band 6, ngVLA Band 6, ngVLA Band 5 and ngVLA Band4, respectively.
The horizontal dotted line represents the luminosity of the star in our model.
}
\label{fig:herbigstars}
\end{center}
\end{figure}

Figure \ref{fig:herbigstars} compares the expected angular separation of the dead-zone inner edge of known 218 Herbig disks with the angular resolution of ALMA and ngVLA obtained in our synthetic images.
The stellar luminosities and distances are taken from \citet{Vioque+18}.
We clearly see that if the angular resolution is 3 mas, which is comparable to the best resolution of ngVLA, the dead-zone inner edge can be spatially resolved for many disks around known Herbig Ae/Be stars.

\begin{figure}[ht]
\begin{center}
\includegraphics[scale=0.73]{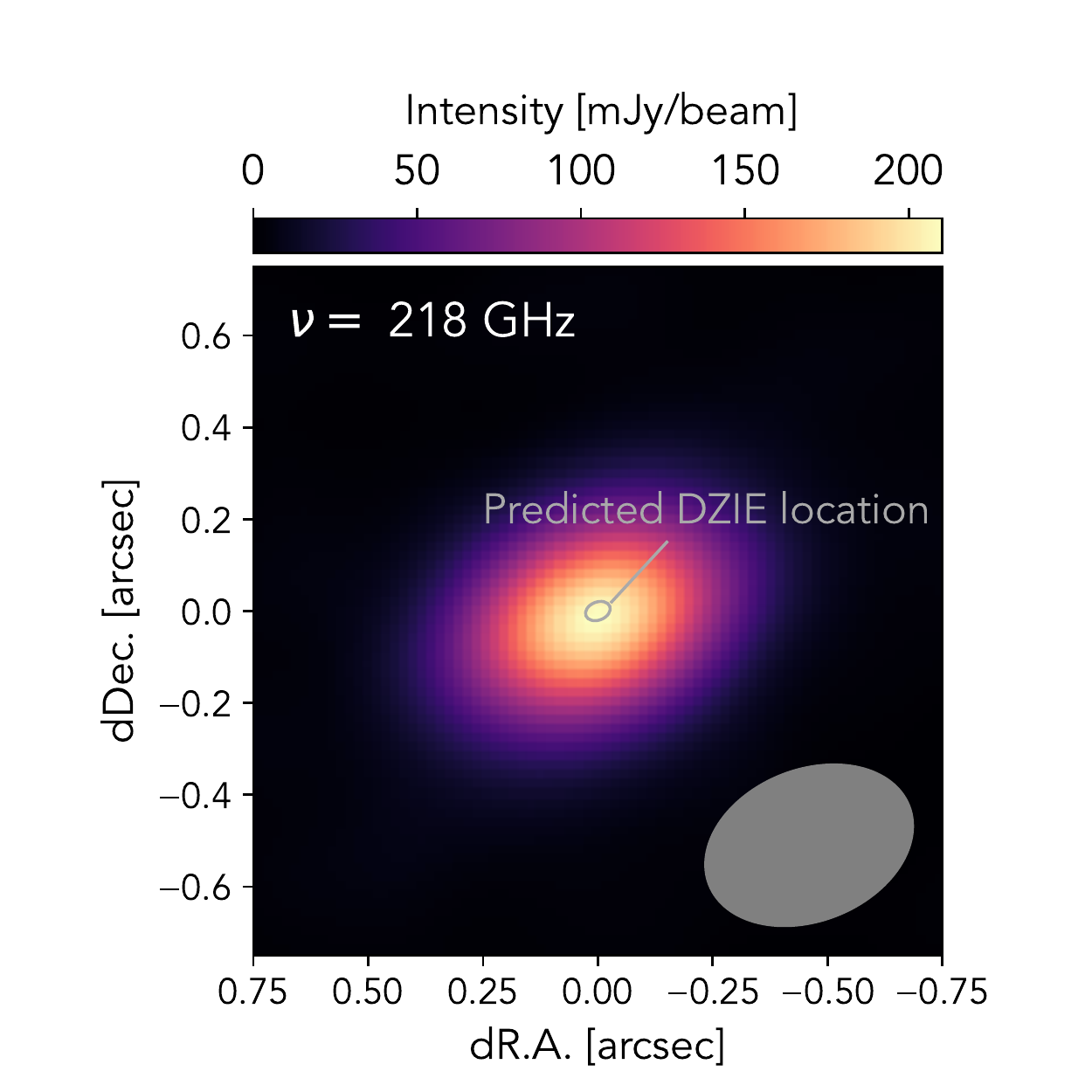}
\caption{
ALMA archival data of the MWC 297 disk observed at Band 6.
The synthesized beam-size is denoted with a gray-filled ellipse at lower right. 
The expected location of the dead-zone inner edge is denoted with a gray-open ellipse.
}
\label{fig:MWC297}
\end{center}
\end{figure}
Even though disks around luminous Herbig Ae/Be stars are expected to be less massive than those around T-Tauri star due to strong irradiation (\citealt{Fuente+03,Alonso-Albi+09}), there are some disks which are found to extend far beyond the dead-zone inner edge.
As an example, the circumstellar disk around MWC 297 would be one of the best targets for probing the dead-zone inner edge.
MWC 297 is a young pre-main-sequence Herbig Be type star with a spectral type of B1.5V, a distance of $\sim$ 375 pc and a luminosity of $\sim39000~L_{\odot}$ \citep{Vioque+18}.
Owing to its high luminosity, the dead-zone inner edge is expected to be located at $\sim$ 21 au ($\sim 56$ mas for $d=375$ pc).
The angular separation between the central star and the dead-zone inner edge can be resolved into $\sim$ 37 beam segments at ngVLA Band 6.
Figure \ref{fig:MWC297} shows the ALMA archival data of the MWC 297 disk observed at ALMA Band 6 (2018.1.00814.S, PI: L. Maud).
Although the angular resolution is not high enough to probe the dead-zone inner edge, the previous observation confirm the dust emission extended to $\sim$ 0.5 arcsec (188 au at $d=375$ pc).
The total flux at 218 GHz ($\lambda=1.38~{\rm \mu m}$) is 273 mJy, significantly brighter than most T-Tauri disks \citep{Andrews+13,Ansdell+18}.
Therefore, the MWC 297 disk would be one of the best targets to explore the dust distribution around the dead-zone inner edge.

It is worth noting that the MWC 297 disk is expected to have a inner disk component which extends inside of the dead-zone inner edge \citep{Guzman+21}. 
It is very unclear if disks around Herbig Ae/Be stars extend inward to near the the dead-zone inner edge because only a small number of Herbig Ae disks has been mapped with high spatial resolution.
From the Spectral Energy Distribution (SED) analysis, some disks
around Herbig Ae/Be stars are expected to have gaps or inner cavities \citep{Maaskant+13}.
Therefore, it is necessary to select potential targets based on their SEDs.

\section{Summary}\label{sec:summary}
We investigated the capabilities of ALMA and ngVLA to detect and spatially resolve the concentration of dust expected by models of dust evolution at the inner edge of the dead-zone around a Herbig star with luminosity of $56~L_{\odot}$.
Herbig Ae/Be stars are useful laboratories for exploring the evolution of rocky grains because of their high luminosity.
We used models of dust evolution around the dead-zone inner edge given by \citet{Ueda+19} and produced predictions for the dust continuum emission for future observations with ALMA and ngVLA.
We found that, thanks to its unprecedented angular resolution and sensitivity, ngVLA can detect and spatially resolve a dust concentration at the dead-zone inner edge, with a reasonable integration time of 10 hrs at $\lambda=3,7$ mm and 1 cm.
The dust concentration is expected to be optically thick at the ALMA wavelength and cannot be spatially resolved due to its limited resolution.
On the other hand, we can potentially infer the dust concentration with current VLA by using the flux density emitted from the innermost region with a diameter of 7 au which is comparable to the highest resolution of VLA with a distance of 140 pc.
Finally, we examined the expected location of the dead-zone inner edge for 218 Herbig Ae/Be stars and found that we can spatially resolve the dead-zone inner edge for tens of those disks with ngVLA resolution.
Particularly, the MWC 297 disk would be one of the best targets because the separation between the central star and the dead-zone inner edge can be resolved into $\sim$ 37 beam segments with ngVLA Band 6.
These observations will allow us to shed light on the process of dust concentration in the inner disk regions, an important process for the formation of rocky planets/planetesimals in young protoplanetary disks.

\acknowledgments
We thank an anonymous referee for useful comments.
T.U. is supported by JSPS KAKENHI Grant Numbers JP19J01929.
Part of this work was supported by NAOJ Overseas Visit Program for Young Researchers, NINS.
M.F. acknowledge funding from the European Research Council (ERC) under the European Union’s Horizon 2020 research and innovation program (grant agreement No. 757957).
This paper makes use of the following ALMA data: ADS/JAO.ALMA$\#2018.1.00814.{\rm S}$.
ALMA is a partnership of ESO (representing its member states), NSF (USA) and NINS (Japan), together with NRC (Canada), MOST and ASIAA (Taiwan), and KASI (Republic of Korea), in cooperation with the Republic of Chile. The Joint ALMA Observatory is operated by ESO, AUI/NRAO and NAOJ.
\software{RADMC-3D \citep{RADMC}}

\appendix
\section{Mid-plane temperature profiles} \label{sec:app}
\begin{figure}[ht]
\begin{center}
\includegraphics[scale=0.38]{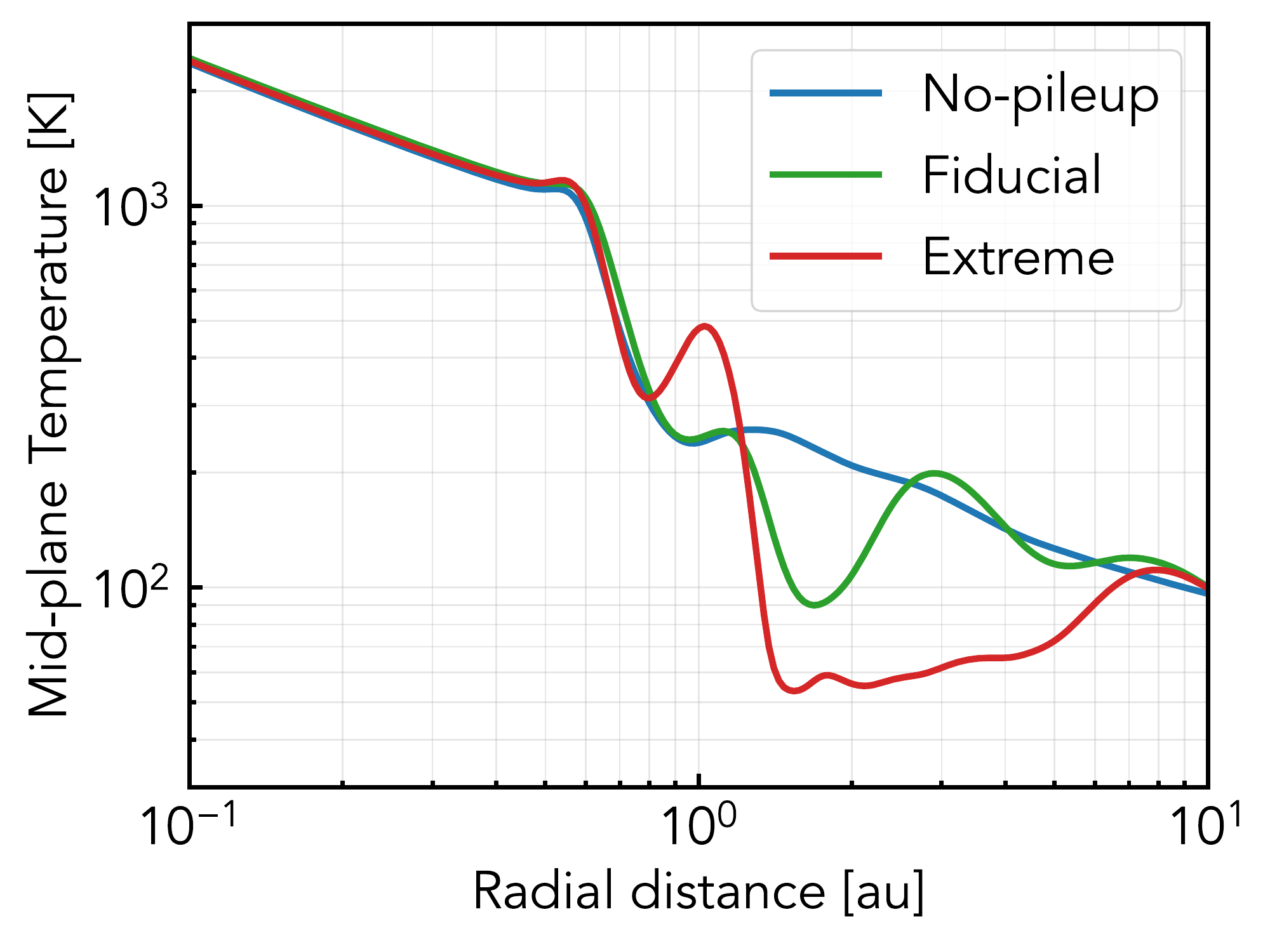}
\caption{
Mid-plane temperature profiles of our models.
}
\label{fig:temp}
\end{center}
\end{figure}

Figure \ref{fig:temp} shows the mid-plane temperature profile obtained from RADMC-3D simulations.
Inside $\sim$ 0.5 au, the temperature follows a temp profile given under the optically thin regime as most of the dust component evaporates.
At $\sim$ 0.5 au, the temperature steeply decreases because the disk becomes optically thick for the stellar irradiation.
For the models with a dust pileup, the dust-pileup cast blocks off the stellar light and cast a shadow just behind it, results in a dip in the temperature \citep{Ueda+19}.
In our models, the temperature between 0.5 and 1 au is lower than that obtained by the simulations that treat a detailed rim structure (e.g., \citealt{Dullemond+01,Flock2016,Flock2017,Ueda+17}). 
The disk structure around the disk rim sensitively depends on the dust sublimation and direct irradiation onto the disk rim. 
The detailed modelling of the disk rim is not trivial because the dust sublimation process and rim structure depends on each other, and will be provided in a future work. 

\bibliographystyle{aasjournal}
\bibliography{reference}

\end{document}